\documentclass[12pt]{article}

\usepackage{amsmath}
\usepackage{amsfonts}
\usepackage{amssymb}
\usepackage{graphicx}
\usepackage{psfrag}
\usepackage{graphicx}
\usepackage{psfrag}

\newcommand{\be}{\begin{equation}}
\newcommand{\ee}{\end{equation}}
\newcommand{\ba}{\begin{eqnarray}}
\newcommand{\ea}{\end{eqnarray}}

\begin{document}
\begin{center}
{\bf
SOLUTION OF SECOND ORDER SUPERSYMMETRICAL INTERTWINING RELATIONS IN MINKOWSKI PLANE}\\
\vspace{1cm}
{\large M. V. Iof\/fe$^{1,}$\footnote{E-mail: m.ioffe@spbu.ru, corresponding author},
E. V. Kolevatova$^{1,}$\footnote{E-mail: e.v.krup@yandex.ru},
D. N. Nishnianidze}$^{2,1,}$\footnote{E-mail: cutaisi@yahoo.com}\\
\vspace{0.5cm}
$^1$ Saint Petersburg State University, 7/9 Universitetskaya nab., St.Petersburg, 199034 Russia.\\
$^2$ Akaki Tsereteli State University, 4600 Kutaisi, Georgia.\\

\end{center}
\vspace{0.5cm}
\hspace*{0.5in}
\vspace{1cm}
\hspace*{0.5in}
\begin{minipage}{5.0in}
{\small
Supersymmetrical (SUSY) intertwining relations are generalized to the case of quantum Hamiltonians in Minkowski space. For intertwining operators (supercharges) of second order in derivatives the intertwined Hamiltonians correspond to completely integrable systems with the symmetry operators of fourth order in momenta. In terms of components, the itertwining relations correspond to the system of nonlinear differential equations which are solvable with the simplest - constant - ansatzes for the "metric" matrix in second order part of the supercharges. The corresponding potentials are built explicitly both for diagonalizable and nondiagonalizable form of "metric" matrices, and their properties are discussed.
}
\end{minipage}


\section{Introduction.}
During last decades, the method of supersymmetry (SUSY) has been successfully used to study different problems
in Quantum Mechanics \cite{ssqm}, \cite{ai}: both traditional ones (the spectral problem for Schr\"odinger equation as example), and relatively new
(the problems of quantum design as example). Actually, this approach gave an essential impetus to the development of Quantum
Mechanics and its applications. Initially, SUSY formalism was developed for different aspects of one-dimensional Quantum Mechanics,
but later on its generalizations for the case of systems in multidimensional spaces were invented \cite{ioffe-reviews}, \cite{ai}. This extension provided an essential achievement of SUSY, since the old analytical methods are quite limited for study such class of systems.

At first, the generalization from the one-dimensional case to the arbitrary dimensionality of space was built by the
most simple way - by means of supercharges of first order in momenta \cite{abei}. Despite a number of interesting applications were found
in this way \cite{pauli}, \cite{calogero}, this version had some disadvantage: the whole construction inevitably contains matrix components of SuperHamiltonian,
and their interpretation is not always obvious. For the lowest (two-dimensional) space dimensionality,
the alternative way to use SUSY ideas was elaborated \cite{ai}, \cite{ioffe-reviews}, \cite{david}. Namely, the standard SUSY algebra with three elements - SuperHamiltonian $\widehat H$ and two supercharges $\widehat Q^{\pm}$ - was generalized as follows:
\ba
[\widehat H, \widehat Q^{\pm}]&=&0; \label{commut}\\
\{\widehat Q^+, \widehat Q^-\}&=&\widehat R; \label{anti}\\
\{\widehat Q^+, \widehat Q^+\}&=&\{\widehat Q^-, \widehat Q^-\}=0, \label{nilpot}
\ea
where $\widehat Q^{\pm}$ are of second order in momenta. Equations (\ref{commut}), (\ref{nilpot}) are kept the same as in standard SUSY Quantum Mechanics \cite{ssqm}, but
the linear function of $\widehat H$ in r.h.s. of (\ref{anti}) is replaced here by some diagonal fourth order operator $\widehat R,$ which generally speaking is not reduced to
a polynomial of $\widehat H.$ The relations (\ref{commut}) correspond to SUSY intertwining relations and they just provide the isospectrality
of components of the SuperHamiltonian. This approach to two-dimensional Quantum Mechanics with operators $\widehat Q^{\pm}$ of second order was developed in a
series of papers \cite{david}. Briefly, the main results of study of two-dimensional systems by means of second order SUSY approach are the following.
Though the general solution of intertwining relations can not be constructed, a wide class of solutions, i.e. of Hamiltonians and supercharges,
was built explicitly \cite{david} due to suitable choice of ansatzes. All these Hamiltonians possess the symmetry operators of fourth order in momenta, thereby being completely integrable. Among these systems three specific models were found \cite{new} such that the corresponding Schr\"odinger equations are solvable. To be more precise, depending on the values of coupling constants,
the Schr\"odinger equation was solved exactly or quasi exactly. In the first case, all wave functions and the whole discrete spectrum were found analytically, while
in the second case - a part of wave functions and energy values were found.

Until now, the multidimensional generalization of SUSY method was used mainly in terms of Cartesian coordinates in the Euclidean space. However,
it seems to be interesting to formulate the SUSY approach both in arbitrary coordinates and beyond the Euclidean space. In particular, this interest is due to importance
of intertwining relations not only for the old problems of conventional Quantum Mechanics, but also for variety of different problems in modern Theoretical and Mathematical Physics (see \cite{1}, as example). As well, the study of Schr\"odinger-like equations in spaces with different kinds of geometry might be interesting in the framework of some models in cosmology and gravity \cite{2}.
The general $d-$dimensional SUSY Quantum Mechanics with first order supercharges was formulated by means of arbitrary curvilinear coordinates \cite{zhu} (see also \cite{mateos}). The straightforward intertwining of scalar Schr\"odinger Hamiltonians by the first order operators was studied for the simplest two-dimensional non-Euclidean spaces in \cite{samani} (see also \cite{kuru}). It was shown in \cite{samani}, that similarly to the Euclidean space such first order intertwining leads to the Hamiltonians which are amenable to conventional separation of variables, i.e. they are reducible to a pair of one-dimensional problems. In the present paper we expand the method of second order interwining \cite{david} to the systems beyond the Euclidean metric: the case of Minkowski plane will be considered. From the very beginning we are mainly interested in models which do not allow the conventional separation of variables (even in $R-$separation form \cite{miller}). As a byproduct of our construction, again the second order SUSY intertwining in Minkowski plane provides us with a symmetry operator of fourth order in momenta, i.e. the constructed models are completely integrable.

The structure of the paper is the following. After formulation of the considered problem in Section 2, the cases with different forms of second order supercharges are studied. The case when the "metric" $g_{nm}$ in supercharges can be diagonalized is investigated in Section 3, separately for $g_{nm}=diag(1,\,-1)$ (Subsection 3.1) and $g_{nm}=diag(1,\,1)$ (Subsection 3.2). The list of potentials which solve the SUSY intertwining relations is obtained, the properties of potentials are discussed. Section 4 contains results for the nondiagonalizable "metric" $g_{nm}:$ a pair of SUSY partners $H^{(1, 2)}$ with suitable properties is built. The final Section 5 includes some conclusions and discussion on applicability of the SUSY separation of variables procedure for several constructed models with $V^{(2)}=const.$

\section{Formulation of the problem.}

We shall consider the class of Hermitian Hamiltonians in two-dimensional space with Minkowski metric:
\be
  H^{(i)}(\vec x) = h_{kl}\partial_k\partial_l + V^{(i)}(\vec x);\quad i=1,2;\quad \partial_l\equiv \partial / \partial x_l; \quad h_{kl}=diag(1,-1), \label{ham}
\ee
which participate in the SUSY intertwining relations:
\begin{equation}\label{intertw}
H^{(1)}Q^+ = Q^+ H^{(2)};\quad Q^-H^{(1)} = H^{(2)}Q^-.
\end{equation}
The Hamiltonians $H^{(1)},\, H^{(2)}$ are the components of $2\times 2$ diagonal matrix SuperHamiltonian $\widehat H,$
and the intertwining operators $Q^{\pm}$ are the components of the off-diagonal matrix operators $\widehat Q^{\pm}$ (see (\ref{commut}) - (\ref{nilpot})). These operators are arbitrary second order differential operators:
\begin{equation}\label{charge}
Q^+ = g_{nm}(\vec x)\partial_n\partial_m + \widetilde C_p(\vec x)\partial_p + B(\vec x);\quad Q^-=(Q^+)^{\dagger}
\end{equation}
with real coefficient functions $\widetilde C_1(\vec x), \,\widetilde C_2(\vec x), \, B(\vec x)$ and symmetric real matrix $g_{n m}(\vec x).$ The intertwining relations (\ref{intertw}) provide
the isospectrality of Hamiltonians $H^{(1)},\,H^{(2)}$ up to zero modes of intertwining operators $Q^{\pm}.$ Both Hamiltonians participating in SUSY intertwining relations
are completely integrable. Indeed, Eqs.(\ref{intertw}) lead immediately to commutation relations:
\be
[H^{(1)},\, R^{(1)}] = [H^{(2)},\, R^{(2)}]=0;\quad R^{(1)}\equiv Q^+Q^-; \quad R^{(2)}\equiv Q^-Q^+, \label{RR}
\ee
where $R^{(1, 2)}$ are the elements of diagonal matrix operator $\widehat R$ in (\ref{anti}). These fourth order operators are the symmetry operators for systems with Hamiltonians $H^{(1, 2)},$ correspondingly.

The SUSY approach in Quantum Mechanics consists of solving intertwining relations (\ref{intertw}), i.e. of finding the potentials $V^{(1, 2)}$ and
corresponding coefficient functions $\widetilde C_1(\vec x), \,\widetilde C_2(\vec x), \, B(\vec x),\, g_{nm}(\vec x).$
In such a general formulation, intertwining relations (\ref{intertw}) can be rewritten as a very complicate system of nonlinear differential equations for
eight functions $g_{nm}(\vec x), V^{(1,2)}(\vec x), \widetilde C_p(\vec x), B(\vec x),$ and it has no chances to be solved in a general form. Therefore, we have to choose some suitable ansatzes to simplify the task.

\section{Solutions for diagonalizable $g_{nm}$}

First of all, the particular form of "metric" $g_{nm}$ in the supercharge will be taken: having a pretty rich experience in conventional two-dimensional Quantum Mechanics,
we restrict ourselves to the constant matrix $g_{nm}.$ At first, we shall consider the constant matrices $g_{nm},$ which can be diagonalized by linear constant transformation of coordinates keeping the form of kinetic term $(\partial_1^2-\partial_2^2)$ of the Hamiltonian unchanged. By making $g_{12}$ a positive, the direct calculations show that matrix $g_{nm}$ can be diagonalized iff it satisfies the condition:
\begin{equation}
g_{11}+g_{22} > 2g_{12} \quad or \quad  g_{11}+g_{22} < -2g_{12},  \nonumber
\nonumber
\end{equation}
and this condition will be assumed fulfilled in the present Section. Therefore, from this point on the matrix $g_{nm}$ will be used in the form:
\begin{equation}\label{g}
g_{nm}=diag(1, -a^2)
\end{equation}
with pure real or pure imaginary parameter $a.$

Intertwining relations (\ref{intertw}) can be represented as the vanishing of a third order
differential operator in partial derivatives. By separating the coefficients of different powers of derivatives, one obtains a system of six nonlinear differential equations for the real functions $V^{(1),(2)}(\vec x),\, \widetilde C_p(\vec x),\, B(\vec x):$
\begin{eqnarray}
&&h_{ik}(\partial_i\widetilde C_p) + h_{ip}(\partial_i\widetilde C_k) + g_{kp}(V^{(1)}-V^{(2)}) = 0; \label{1} \\
&&h_{ik}(\partial_i\partial_k\widetilde C_p) + h_{ip}(\partial_iB) + \widetilde C_p(V^{(1)}-V^{(2)}) -2g_{mp}(\partial_mV^{(2)}) = 0; \label{2} \\
&&h_{ik}(\partial_i\partial_kB) + B(V^{(1)}-V^{(2)}) -\widetilde C_p(\partial_pV^{(2)})-g_{mn}(\partial_n\partial_mV^{(2)}) = 0, \label{3}
\end{eqnarray}
where the metrics $h_{ik}$ and $g_{nm}$ of the form (\ref{ham}) and (\ref{g}) must be used. It is convenient to define:
\begin{equation}\label{V}
V^{(2)}-V^{(1)} \equiv 2V.
\end{equation}

Starting from three equations contained in Eq.(\ref{1}), one obtains:
\be
\partial_1\widetilde C_1=V; \quad \partial_2\widetilde C_2=a^2V; \quad \partial_1\widetilde C_2=\partial_2\widetilde C_1 . \label{2-4-3}
\ee
It is reasonable to write down solutions of (\ref{2-4-3}) in new coordinates
\begin{equation}
y_1\equiv x_1-ax_2;\quad y_2\equiv x_1+ax_2, \nonumber
\end{equation}
which allow to express solutions in terms of new arbitrary functions $C_1(y_1)$ and $C_2(y_2):$
\be
\widetilde C_1 = C_1(y_1)+C_2(y_2); \quad \widetilde C_2 = a\biggl(C_2(y_2)-C_1(y_1)\biggr);\quad V = C_1^{\prime}(y_1)+C_2^{\prime}(y_2) \label{3-4-3}
\ee
($C_i^{\prime}$ means differentiation with respect to its argument).

In its turn, two equations contained in Eq.(\ref{2}) are ($\partial_i$ still mean derivatives over $x_i$):
\ba
&&(\partial_1^2-\partial_2^2)\widetilde C_1(y_1) + 2 \partial_1B(\vec y) - 2V(\vec y)\widetilde C_1(y_1) - 2\partial_1V^{(2)}(\vec y)=0; \label{a}\\
&&(\partial_1^2-\partial_2^2)\widetilde C_2(y_2) - 2 \partial_2 B(\vec y) - 2V(\vec y)\widetilde C_2(y_2) + 2a^2\partial_2V^{(2)}(\vec y)=0. \label{b}
\ea
Adding and subtracting these equations, one obtains:
\ba
(1-a^2)C_2^{\prime\prime}(y_2)&+&2\partial_{y_1}B(\vec y)-2C_2(y_2)\biggl(C_1^{\prime}(y_1)+C_2^{\prime}(y_2)\biggr)-\nonumber\\
&-&\biggl[(1+a^2)\partial_{y_1}+(1-a^2)\partial_{y_2}\biggr]V^{(2)}(\vec y)=0; \label{3-star}\\
(1-a^2)C_1^{\prime\prime}(y_1)&+&2\partial_{y_2}B(\vec y)-2C_1(y_1)\biggl(C_1^{\prime}(y_1)+C_2^{\prime}(y_2)\biggr)-\nonumber\\
&-&\biggl[(1-a^2)\partial_{y_1}+(1+a^2)\partial_{y_2}\biggr]V^{(2)}(\vec y)=0. \label{4-star}
\ea
From the condition of compatibility of these equations (if $a^2\neq 1$), one derives the general expression for potential $V^{(2)}(\vec y)$:
\begin{equation}\label{V_2}
V^{(2)}(\vec y)=C_1^{\prime}(y_1)+C_2^{\prime}(y_2)-\frac{1}{1-a^2}\biggl(C_1^2(y_1)+C_2^2(y_2)\biggr)+F_+(y_1+y_2)+F_-(y_1-y_2),
\end{equation}
where $F_+,\, F_-$ are arbitrary functions of $y_{\pm}\equiv y_1\pm y_2,$ respectively. Substitution of (\ref{V_2}) into (\ref{3-star}) and (\ref{4-star}) gives the general expression for
$B(\vec y):$
\begin{equation}
B(\vec y)=C_1(y_1)C_2(y_2)+\frac{1}{2}(1+a^2)V^{(2)}(\vec y)+\frac{1}{2}(1-a^2)\biggl(F_+(y_1+y_2)-F_-(y_1-y_2)\biggr)+\beta , \nonumber
\end{equation}
where $V^{(2)}$ is given by (\ref{V_2}), and $\beta$ is an arbitrary real constant.

Thus, to finish solving of intertwining relations (\ref{intertw}) the last equation - Eq.(\ref{3}) - has to be solved. It is useful to calculate an additional
derivative of Eq.(\ref{2}) and to substitute it into (\ref{3}):
\begin{equation}\label{6-1}
-\frac{1-a^4}{2}\biggl(C_1^{\prime\prime\prime}(y_1)+C_2^{\prime\prime\prime}(y_2)\biggr) + (1+a^2)\biggl(V(\vec y)\biggr)^2 -\widetilde C_p\partial_p\biggl(V^{(2)}(\vec y)-V(\vec y)\biggr)-2VB=0.
\end{equation}
After a series of manipulations, Eq.(\ref{3}) can be transformed to the form
of functional-differential equation for functions $C_1(y_1),\,C_2(y_2),\,F_+(y_1+y_2),\, F_-(y_1-y_2):$
\ba
&&\partial_{y_1}\Biggl[\frac{1-a^4}{2}C_1^{\prime\prime} - \frac{1+a^2}{1-a^2}\biggl(C_1^3+C_1C_2^2\biggr)+2\beta C_1+2C_1\biggl(F_++a^2F_-\biggr)\Biggr]=\nonumber\\
&&=-\partial_{y_2}\Biggl[\frac{1-a^4}{2}C_2^{\prime\prime} - \frac{1+a^2}{1-a^2}\biggl(C_2^3+C_2C_1^2\biggr)+2\beta C_2+2C_2\biggl(F_++a^2F_-\biggr)\Biggr]. \label{last}
\ea

\subsection{The case $a^2=-1$}

This equation seems difficult to solve for an arbitrary value of the parameter $a.$ But it is essentially simplified for $a=-i,$ i.e. for $g_{nm}=diag(1,1),$ since several terms in (\ref{last}) vanish in this case:
\be
\partial_{y_1}\Biggl[C_1(y_1)\biggl(F_+(y_1+y_2)-F_-(y_1-y_2)\biggr)\Biggr]=-\partial_{y_2}\Biggl[C_2(y_2)\biggl(F_+(y_1+y_2)-F_-(y_1-y_2)\biggr)\Biggr], \label{last-1}
\ee
the constant $\beta$ being absorbed by terms $(F_+ - F_-)$. The choice $a=-i$ means that we deal with complex variables $y_1,\, y_2,$ which are mutually conjugated $y_1=\bar y_2.$ Correspondingly, the conditions that $\widetilde C_1(y_1),\, \widetilde C_2(y_2),\,F_+(y_1+y_2),\, F_-(y_1-y_2)$ are real valued impose the following restrictions:
\begin{equation}\label{reality}
C_1(y_1)=\overline{C_2(y_2)}; \quad F_+(y_1+y_2)=real;\quad F_-(y_1-y_2)=real,
\end{equation}
where the bar means complex conjugation.

The equation (\ref{last-1}) is a functional-differential one, and therefore, some nontrivial procedure must be applied to solve it. Fortunately, this work was already done: an analogous equation was investigated in detail in a different context in papers \cite{david}, in reviews \cite{ioffe-reviews}, \cite{ai} and in a compact form in \cite{bin}.
Specifically, the intertwining relations were studied there for Hamiltonians in Euclidean plane with Lorentz-like form $(\partial_1^2-\partial_2^2)$ of second order terms in supercharges. After
considering some simplifying ansatzes \cite{david}, the general solution of the problem was finally formulated in \cite{bin}.
The relation between Eq.(\ref{last-1}) and equations (BIN-6), (BIN-7) of \cite{bin} (here and below arbitrary equation (n) of \cite{bin} is denoted as (BIN-n)) is established simply by identifying our variables $y_{1,2}$ with $x_{\pm}$
of that papers, correspondingly. Thus, all solutions obtained in \cite{bin}, \cite{david}, \cite{ioffe-reviews}, \cite{ai} must be passed through the filter of (\ref{reality}), and the result will represent the required solutions of (\ref{last-1}).
After checking of all possible variants from \cite{ai} (Subsection 8.2.1 with reducible and Subsection 8.2.2 with irreducible supercharges) and from \cite{bin} (Section 3),
the full list of suitable potentials can be obtained.

The first solution can be built for the case of factorizable $(F_+-F_-),$ which corresponds to variants I and II of \cite{bin}:
\begin{equation}
F_+(2x_1)-F_-(2ix_2)=\Phi(y_1)\overline{\Phi}(y_2);\quad  y_1=x_1+ix_2;\quad y_2=\bar{y}_1=x_1-ix_2. \nonumber
\end{equation}
Indeed, for such factorization, function $\Phi$ can be found explicitly:
\begin{equation}
\frac{\Phi^{\prime\prime}(y_1)}{\Phi(y_1)}=\frac{\overline{\Phi}^{\prime\prime}(y_2)}{\overline{\Phi}(y_2)}=\lambda^2;\quad \Phi(y)\sim \cosh(\lambda y); \nonumber
\end{equation}
with real constant $\lambda^2,$ Obviously, it allows to solve (\ref{last-1}):
\begin{equation}\label{1111}
C_1(y_1)=\overline{C_2(y_2)}=i\mu \frac{\Phi^{\prime}(y_1)}{\Phi(y_1)}+\frac{2\nu}{\Phi(y_1)};\,\, F_+(2x_1)=k\cosh(2\lambda x_1);\, F_-(2ix_2)=-k\cos(2\lambda x_2),
\end{equation}
with arbitrary real constants $k,\,\mu$ and complex constant $\nu ,$ leading to the real partner potentials written in terms of initial coordinates $x_1,\, x_2:$
\ba
&&V^{(1,2)}(\vec x)= \biggl(\cosh^2(\lambda x_1)-\sin^2(\lambda x_2)\biggr)^{-2}\Biggl[(\lambda^2\mu^2+\nu_1^2-\nu_2^2)\biggl(\sinh^2(\lambda x_1)\sin^2(\lambda x_2)-\nonumber\\
&&-\cosh^2(\lambda x_1)\cos^2(\lambda x_2)\biggr)-(\nu_1\nu_2 \pm \mu\lambda^2)\sinh(2\lambda x_1)\sin(2\lambda x_2)\Biggr]+\nonumber\\
&&+ 2\lambda \biggl(\cosh^2(\lambda x_1)-\sin^2(\lambda x_2)\biggr)^{-1}\Biggl[(\mu\nu_2\mp \nu_1)\sinh(\lambda x_1)\cos(\lambda x_2)-\nonumber\\
&&-(\mu\nu_1\mp \nu_2)\cosh(\lambda x_1)\sin(\lambda x_2)\Biggr]+k\biggl(\cosh(2\lambda x_1)-\cos(2\lambda x_2)\biggr), \label{pot1}
\ea
(here and below all potentials are defined up to an arbitrary real shift of energy). These potentials have singularities in the points $x_1=0,\, x_2=\frac{\pi}{2\lambda}(2n+1).$ Expressions (\ref{pot1}) are rather cumbersome but one must remember that we are free to choose some of constants to be zero.

The case VII of the paper \cite{bin} is described in Eq.(BIN-18) of that paper.
It corresponds to three different options. The first is:
\begin{equation}
C_{1,2}(y_{1,2})=\frac{k}{\sinh(\lambda y_{1,2})};\,
F_{+}(2x_{1})=n \cosh(2\lambda x_{1})+\frac{m}{\sinh^2(\lambda x_{1})};\, F_-(2ix_2)=-n \cos(2\lambda x_2)+\frac{m}{\sin^2(\lambda x_2)}
\nonumber
\end{equation}
($k,\,n,\,m$ - real constants). The corresponding potentials are:
\ba
&&V^{(1,2)}(\vec x)= \biggl(\cosh(2\lambda x_1)-\cos(2\lambda x_2)\biggr)^{-2}\Biggl[\pm 8k\lambda \cosh(\lambda x_1)\cos(\lambda x_2)\biggl(\sinh^2(\lambda x_1)-\sin^2(\lambda x_2)\biggr)-\nonumber\\
&&- k^2\biggl(8\sinh^2(\lambda x_1)\cos^2(\lambda x_2)-2\biggl(\cosh(2\lambda x_1)-\cos(2\lambda x_2)\biggr)\biggr)\Biggr] + n\biggl(\cosh(2\lambda x_1)+\cos(2\lambda x_2)\biggr)+\nonumber\\
&&+m\biggl(\sinh^{-2}(\lambda x_1)-\sin^{-2}(\lambda x_2)\biggr) ,  \label{2222}
\ea
being singular along the coordinate axes. These singularities - besides ones at points $x_1=0,\, x_2=\frac{\pi}{\lambda}k$ - can be removed by the choice $m=0.$

The second option gives the potentials of similar form:
\begin{equation}
C_{1,2}(y_{1,2})=\frac{k}{\cosh(\lambda y_{1,2})};\,
F_{+}(2x_1)=n \cosh(2\lambda x_1)+\frac{m}{\sinh^2(\lambda x_1)};\, F_-(2ix_2)=-n \cos(2\lambda x_2)+\frac{m}{\cos^2(\lambda x_2)}  \nonumber
\end{equation}
\ba
&&V^{(1,2)}(\vec x)= \biggl(\cosh(2\lambda x_1)+\cos(2\lambda x_2)\biggr)^{-2}\Biggl[\pm 8k\lambda \sinh(\lambda x_1)\cos(\lambda x_2)\biggl(\cosh^2(\lambda x_1)+\sin^2(\lambda x_2)\biggr)-\nonumber\\
&&- k^2\biggl(8\cosh^2(\lambda x_1)\cos^2(\lambda x_2)-2(\cosh(2\lambda x_1)+\cos(2\lambda x_2))\biggr)\Biggr] + n\biggl(\cosh(2\lambda x_1)-\cos(2\lambda x_2)\biggr)+\nonumber\\
&&+m\biggl(\sinh^{-2}(\lambda x_1)+\cos^{-2}(\lambda x_2)\biggr),  \label{22222222}
\ea
again with removable singularities along both axes, but non-removable at the points $x_1=0,\, x_2=\frac{\pi}{2\lambda}(2l+1).$

The potentials of third option of the case VII correspond to:
$$C_1(y_1)= b/y_1,\quad C_2(y_2)= b/y_2, \quad F_+(2x_1)=nx_1^2+mx_1^{-2},\quad F_-(2ix_2)=-nx_2^2-mx_2^{-2},$$
and potentials are:
\be
V^{(1,2)}(\vec x)=-b(b\mp 2)(x_1^2+x_2^2)^{-2}(x_1^2-x_2^2)+n(x_1^2-x_2^2)+m(x_1^{-2}-x_2^{-2}). \label{2222222}
\ee
They are singular along the axes (if $m\neq 0$), their unlimited decrease at infinity for $\mid x_2\mid > \mid x_1 \mid$ can be avoided by taking $n=0.$
The potentials (\ref{2222222}) obey the property of shape invariance:
\be
V^{(1)}(x_1, x_2; b)=V^{(2)}(x_1, x_2; b-2). \label{shape}
\ee
Initially, this property was introduced \cite{genden} in the framework of one-dimensional SUSY Quantum Mechanics. Later on it was generalized \cite{new}, \cite{ioffe-reviews}
for the two-dimensional case. In particular, if some "basic" wave functions $\Psi^{(1)}_N(\vec x)$ of $H^{(1)}$ are known, shape invariance provides the construction of the variety of other wave functions $\Psi^{(1)}_{N, k}$ by means of the chain of operators $Q^+$ with different values of $b$ (see details of the method for two-dimensional models in \cite{new}, \cite{ioffe-reviews}). The "basic" wave functions of $H^{(1)}$ can be calculated, for example, for parameters $b=n=0,$ since the variables are separated in this case, and the problem is reduced to two one-dimensional equations with simple potential $m/x^2.$

The conventional separation of variables in $H^{(1)}$ for $b=2,\, n=0$ allows to apply also one of two variants of SUSY-separation of variables \cite{new}, \cite{ioffe-reviews}. For the mentioned values of parameters, wave functions of $H^{(1)}$ can be calculated straightforwardly by the procedure of standard separation. After that, wave functions of $H^{(2)}$ for the same values of parameters can be found immediately by action of $Q^-$ according to intertwining relations (\ref{intertw}). This is realization of well known quasi-isospectrality of superpartner Hamiltonians in SUSY Quantum Mechanics \cite{ssqm} in a two-dimensional context.


The case VI in Eq.(BIN-18) of \cite{bin} corresponds to:
\begin{equation}
C_1(y_1)=\frac{ia}{y_1};\quad C_2(y_2)=-\frac{ia}{y_2}; \quad F_+(2x_1)=cx_1^2+bx_1^4;\quad F_-(2ix_2)=-cx_2^2+bx_2^4,  \nonumber
\end{equation}
and partner potentials are:
\begin{equation}\label{333}
V^{(1,2)}(\vec x)=\pm 4a x_1x_2(x_1^2+x_2^2)^{-2}+a^2(x_1^2-x_2^2)(x_1^2+x_2^2)^{-2}+c(x_1^2-x_2^2)+b(x_1^4+x_2^4).
\end{equation}
They are singular at the origin, and their behaviour at infinity depends on the positivity of $b.$

Analogously, the potentials corresponding to cases III and IV of \cite{bin} can be calculated. They give four different pairs of
partner potentials:
\ba
&&V^{(1,2)}= \biggl(\cosh(2\lambda x_1)-\cos(2\lambda x_2)\biggr)^{-2}\Biggl[8c(-b\pm \lambda)\cosh(\lambda x_1)\cos(\lambda x_2)\biggl(\sinh^2(\lambda x_1)-\sin^2(\lambda x_2)\biggr)-\nonumber\\
&&-8(b^2+c^2\mp 2b\lambda )\biggl(\sinh^2(\lambda x_1)\cos^2(\lambda x_2)-\frac{1}{4}\biggl(\cosh(2\lambda x_1)-\cos(2\lambda x_2)\biggl)\biggl)\Biggr]-\nonumber\\
&&-2a(b\pm \lambda )\cosh(\lambda x_1)\cos(\lambda x_2)
-\frac{a^2}{2}\cosh(2\lambda x_1)\cos(2\lambda x_2)+k\biggl(\sinh^{-2}(\lambda x_1)-\sin^{-2}(\lambda x_2)\biggr) . \label{44}
\ea

\ba
&&V^{(1,2)}= \biggl(\cosh(2\lambda x_1)+\cos(2\lambda x_2)\biggr)^{-2}\Biggl[8c(-b\pm \lambda)\sinh(\lambda x_1)\cos(\lambda x_2)\biggl(\cosh^2(\lambda x_1)+\sin^2(\lambda x_2)\biggr)-\nonumber\\
&&-8(-b^2+c^2\pm 2b\lambda )\biggl(\cosh^2(\lambda x_1)\cos^2(\lambda x_2)-\frac{1}{4}\biggl(\cosh(2\lambda x_1)+\cos(2\lambda x_2)\biggr)\biggr)\Biggr]-\nonumber\\
&&-2a(b\pm \lambda )\sinh(\lambda x_1)\cos(\lambda x_2)
-\frac{a^2}{2}\cosh(2\lambda x_1)\cos(2\lambda x_2)+k\biggl(\cosh^{-2}(\lambda x_1)+\sin^{-2}(\lambda x_2)\biggr) . \label{444}
\ea

\ba
&&V^{(1,2)}= \biggl(\cosh(2\lambda x_1)-\cos(2\lambda x_2)\biggr)^{-2}\Biggl[8c(-b\pm \lambda)\sinh(\lambda x_1)\sin(\lambda x_2)\biggl(\cosh^2(\lambda x_1)+\cos^2(\lambda x_2)\biggr)-\nonumber\\
&&-8(b^2-c^2\mp 2b\lambda )\biggl(\sinh^2(\lambda x_1)\cos^2(\lambda x_2)-\frac{1}{4}\biggl(\cosh(2\lambda x_1)-\cos(2\lambda x_2)\biggl)\biggl)\Biggr]+\nonumber\\
&&+2a(b\pm \lambda )\sinh(\lambda x_1)\sin(\lambda x_2)
+\frac{a^2}{2}\cosh(2\lambda x_1)\cos(2\lambda x_2)+k\biggl(\cosh^{-2}(\lambda x_1)+\cos^{-2}(\lambda x_2)\biggr) . \label{4444}
\ea

\ba
&&V^{(1,2)}= \biggl(\cosh(2\lambda x_1)+\cos(2\lambda x_2)\biggr)^{-2}\Biggl[-8c(-b\pm \lambda)\cosh(\lambda x_1)\sin(\lambda x_2)\biggl(\cos^2(\lambda x_2)-\sinh^2(\lambda x_1)\biggr)+\nonumber\\
&&+8(b^2+c^2\mp 2b\lambda )\biggl(\cosh^2(\lambda x_1)\cos^2(\lambda x_2)-\frac{1}{4}\biggl(\cosh(2\lambda x_1)+\cos(2\lambda x_2)\biggr)\biggr)\Biggr]+\nonumber\\
&&+2a(b\pm \lambda )\sinh(\lambda x_1)\sin(\lambda x_2)
+\frac{a^2}{2}\cosh(2\lambda x_1)\cos(2\lambda x_2)+k\biggl(\sinh^{-2}(\lambda x_1)-\cos^{-2}(\lambda x_2)\biggr) . \label{44444}
\ea
These four potentials have common properties: removable singularities along both axes, and non-removable at the discrete points along the axis $x_1=0.$
For particular choice of parameters, potentials (\ref{44}) - (\ref{44444}) are shape invariant: the first and the last for two cases $a=c=0$ or $b=c,\, a=0,$
but (\ref{44}) and (\ref{44444}) - for $a=c=0,$ only.

The case V of the paper \cite{bin} gives a few additional solutions (we skip again the straightforward calculations):
\ba
&&V^{(1,2)}= -8k(k\mp \lambda)\biggl(\cosh(\lambda x_1)-\cos(\lambda x_2)\biggr)^{-2}\Biggl[\sinh^2(\frac{1}{2}\lambda x_1)\cos^2(\frac{1}{2}\lambda x_2)-
\nonumber\\
&&-\frac{1}{4}\biggl(\cosh(\lambda x_1)-\cos(\lambda x_2)\biggr)\Biggr]
+k_1\biggl(\sinh^{-2}(\lambda x_1)-\sin^{-2}(\lambda x_2)\biggr)+\nonumber\\
&&+k_2\biggl(\frac{\cosh(\lambda x_1)}{\sinh^2(\lambda x_1)}-\frac{\cos(\lambda x_2)}{\sin^2(\lambda x_2)}\biggr) . \label{55}
\ea

\ba
&&V^{(1,2)}= -8k(k\pm \lambda)\biggl(\cosh(\lambda x_1)+\cos(\lambda x_2)\biggr)^{-2}\Biggl[\cosh^2(\frac{1}{2}\lambda x_1)\cos^2(\frac{1}{2}\lambda x_2)-
\nonumber\\ &&-\frac{1}{4}\biggl(\cosh(\lambda x_1)+\cos(\lambda x_2)\biggr)\Biggr]
+k_1\biggl(\sinh^{-2}(\lambda x_1)-\sin^{-2}(\lambda x_2)\biggr)+\nonumber\\
&&+k_2\biggl(\frac{\cosh(\lambda x_1)}{\sinh^2(\lambda x_1)}-\frac{\cos(\lambda x_2)}{\sin^2(\lambda x_2)}\biggr) . \label{555}
\ea

\ba
&&V^{(1,2)}= \biggl(\cosh(\lambda x_1)+\cos(\lambda x_2)\biggr)^{-2}\Biggl[\pm 2k\lambda \sinh(\lambda x_1)\sin(\lambda x_2)-
8k^2 \cosh^2(\frac{1}{2}\lambda x_1)\cos^2(\frac{1}{2}\lambda x_2)+\nonumber\\
&&+2k^2\biggl(\cosh(\lambda x_1)+\cos(\lambda x_2)\biggr)\Biggr]+ k_1\biggl(\cosh(\lambda x_1)+\cos(\lambda x_2)\biggr)+\nonumber\\
&&+k_2\biggl(\cosh(2\lambda x_1)+\cos(2\lambda x_2)\biggr). \label{5555}
\ea

\ba
&&V^{(1,2)}= \biggl(\cosh(\lambda x_1)-\cos(\lambda x_2)\biggr)^{-2}\Biggl[\pm 2k\lambda \sinh(\lambda x_1)\sin(\lambda x_2)-
8k^2 \sinh^2(\frac{1}{2}\lambda x_1)\cos^2(\frac{1}{2}\lambda x_2)+\nonumber\\
&&+2k^2\biggl(\cosh(\lambda x_1)-\cos(\lambda x_2)\biggr)\Biggr]+ k_1\biggl(\cosh(\lambda x_1)-\cos(\lambda x_2)\biggr)+\nonumber\\
&&+k_2\biggl(\cosh(2\lambda x_1)+\cos(2\lambda x_2)\biggr). \label{55555}
\ea

The potentials (\ref{55}) and (\ref{555}) are shape invariant: $V^{(1)}(\vec x;\, k+\lambda )=V^{(2)}(\vec x;\, k )$ for (\ref{55}) and
$V^{(1)}(\vec x;\, k-\lambda )=V^{(2)}(\vec x;\, k )$ for (\ref{555}). For these systems, both procedures which were described above with respect to
potentials (\ref{2222222}) can be applied here as well, although potentials are more complicate. In Figure 1, the potential $V^{(1)}$ of (\ref{44}) for the constants
$k_1=k_2=0,\, k=1,\, \lambda =2$ is presented for illustration. The obvious properties of this potential are: the periodicity in $x_2,$ singularities at $\bigl(x_1=0,\,x_2=n\pi, \, (n=0,\pm 1, \pm 2,...)\bigr)$ and the invariance under reflections $x_1\to -x_1$ or $x_2\to -x_2.$

\vspace{6pt}

\begin{center}
\includegraphics[height=5cm]{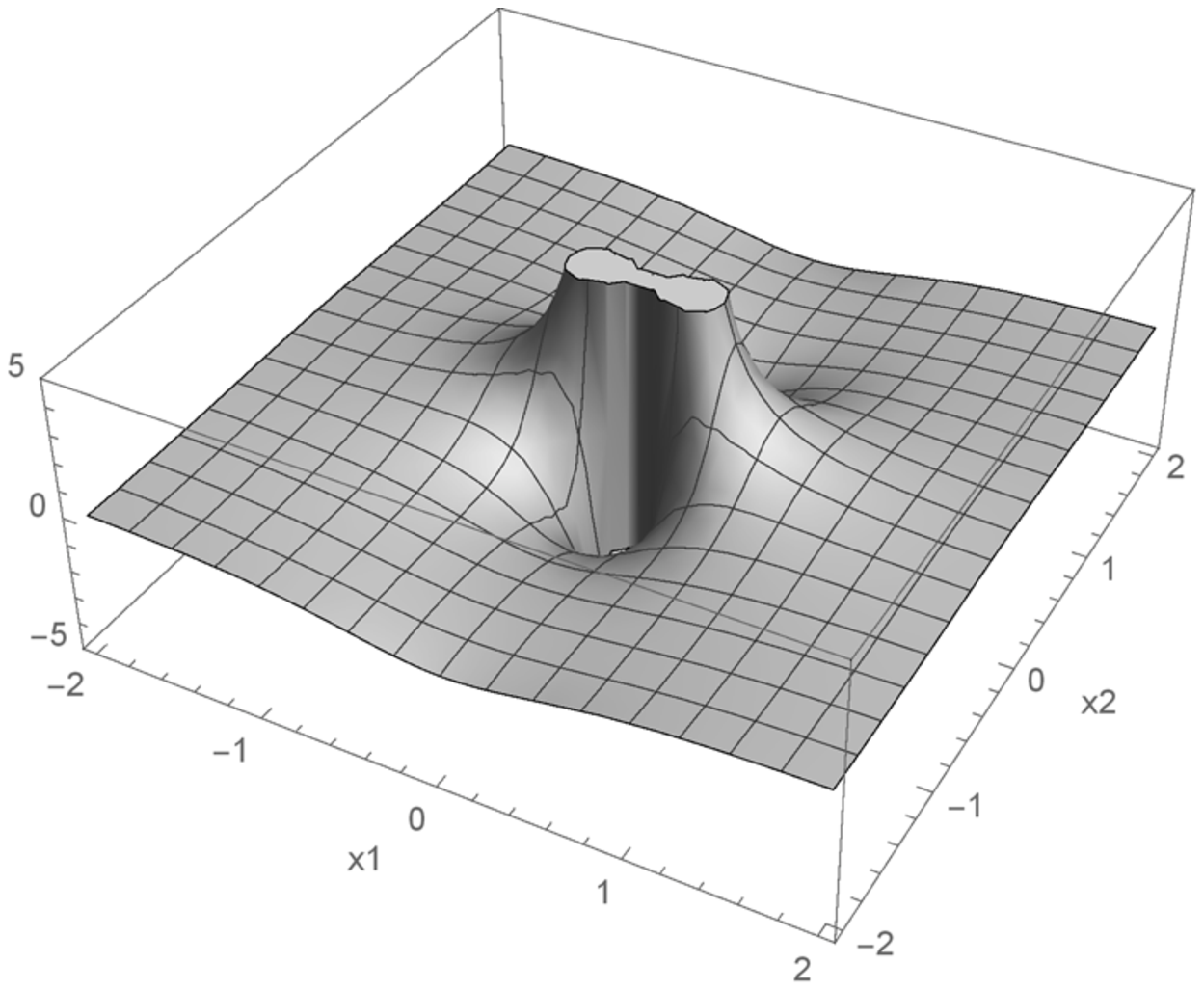}

\noindent{\it Fig.1} Plot of the potential $V^{(1)}$ (\ref{44}) for $ k_1=k_2=0,\, k=1,\, \lambda =2.$

\end{center}

\vspace{4pt}

\subsection{The case $a^2=+1$}

The next particular case - with $a=1$ - will be considered now. The variables $y_i$ are:
\begin{equation}
y_1=x_1-x_2;\quad y_2=x_1+x_2,   \nonumber
\end{equation}
equations (\ref{3-star}), (\ref{4-star}) become:
\ba
2\partial_{y_1}\biggl(B(\vec y)-V^{(2)}(\vec y)\biggr)&=&2C_2(y_2)C_1^{\prime}(y_1)+\biggl(C_2^2(y_2)\biggr)^{\prime}; \label{a1}\\
2\partial_{y_2}\biggl(B(\vec y)-V^{(2)}(\vec y)\biggr)&=&2C_1(y_1)C_2^{\prime}(y_2)+\biggl(C_1^2(y_1)\biggr)^{\prime}, \label{a2}
\ea
and therefore, both $C_1^2(y_1)$ and $C_2^2(y_2)$ are second order polynomials:
\begin{equation}\label{second}
C_1^2(y_1)=\alpha y_1^2+\alpha_1y_1+\beta_1;\quad C_2^2(y_2)=\alpha y_2^2+\alpha_2y_2+\beta_2.
\end{equation}
Eqs.(\ref{a1}), (\ref{a2}) can be integrated:
\ba
2\biggl(B(\vec y)-V^{(2)}(\vec y)\biggr)&=&2C_2(y_2)C_1(y_1)+\biggl(C_2^2(y_2)\biggr)^{\prime}y_1+f_2(y_2);
\nonumber\\
2\biggl(B(\vec y)-V^{(2)}(\vec y)\biggr)&=&2C_1(y_1)C_2(y_2)+\biggl(C_1^2(y_1)\biggr)^{\prime}y_2+f_1(y_1),    \nonumber
\ea
leading to the explicit form of function $f_i:$
\be
f_1(y_1)=\alpha_2y_1 + const;\quad f_2(y_2)=\alpha_1y_2 + const;\label{f}
\ee
and to:
\be
B(\vec y)-V^{(2)}(\vec y)=C_1(y_1)C_2(y_2)+\alpha y_1y_2+\frac{1}{2}(\alpha_2y_1+\alpha_1y_2)+\widetilde\beta \label{V2}
\ee
($\widetilde\beta$ is a constant).

The latter equation allows to express $V^{(2)}$ in terms of functions $B$ and $C_i$ and to substitute it into Eq.(\ref{6-1})
which in our case $a=1$ takes the form:
\begin{equation}
2\biggl(V(\vec y)\biggr)^2-\widetilde C_p\partial_p\biggl(V^{(2)}(\vec y)-V(\vec y)\biggr) -2V(\vec y)B(\vec y)=0.   \nonumber
\end{equation}
From relations (\ref{3-4-3}), (\ref{second}), (\ref{f}), one obtains the following equation for the function $B:$
\begin{equation}\label{BB}
\biggl(C_1\partial_{y_1}+C_2\partial_{y_2}\biggr)\biggl(C_1C_2 B\biggr)=C_1C_2\biggl[2\alpha +2C_1^{\prime}C_2^{\prime}+2\alpha (y_1C_2+y_2C_1)+\alpha_1C_2+\alpha_2C_1\biggr].
\end{equation}
New variables are useful:
\begin{equation}\label{tau}
\tau_1\equiv \int\frac{dy_1}{C_1}+\int\frac{dy_2}{C_2};\quad \tau_2\equiv \int\frac{dy_1}{C_1}-\int\frac{dy_2}{C_2},
\end{equation}
so that the general solution of (\ref{BB}) can be represented as:
\begin{equation}\label{BG}
B=\frac{1}{2C_1C_2}\int \biggl[2\alpha +2C_1^{\prime}C_2^{\prime}+2\alpha (y_1C_2+y_2C_1)+\alpha_1C_2+\alpha_2C_1\biggr]d\tau_1 +\frac{F(\tau_2)}{C_1C_2},
\end{equation}
with an arbitrary function $F(\tau_2).$

Expressions (\ref{second}) provide that functions $C_1,\,C_2$ are real iff $\alpha > 0.$ A suitable translations of $y_1,\, y_2$ allow to fix $\alpha_1= \alpha_2=0,$
so that:
\begin{equation}
C_1^2=\alpha (y_1^2+\omega_1^2);\quad C_2^2=\alpha (y_2^2+\omega_2^2)   \nonumber
\end{equation}
with real constants $\omega_i.$ Thus, up to additive constants:
\begin{equation}
\tau_1=\frac{1}{\sqrt{\alpha}}\ln\biggl[\biggl(y_1+\sqrt{y_1^2+\omega_1^2}\biggr)\biggl(y_2+\sqrt{y_2^2+\omega_2^2}\biggr)\biggr];\quad
\tau_2=\frac{1}{\sqrt{\alpha}}\ln\Biggl[\frac{(y_1+\sqrt{y_1^2+\omega_1^2})}{(y_2+\sqrt{y_2^2+\omega_2^2})}\Biggr];  \nonumber
\end{equation}
and
\ba
C_1&=&\frac{\sqrt{\alpha}}{2\gamma_1}\Biggl(\gamma_1^2\exp\biggl(\frac{\sqrt{\alpha}}{2}(\tau_1+\tau_2)\biggr)+
\omega_1^2\exp\biggl(-\frac{\sqrt{\alpha}}{2}(\tau_1+\tau_2)\biggr)\Biggr);
\nonumber\\
C_2&=&\frac{\sqrt{\alpha}}{2\gamma_2}\Biggl(\gamma_2^2\exp\biggl(\frac{\sqrt{\alpha}}{2}(\tau_1-\tau_2)\biggr)+
\omega_2^2\exp\biggl(-\frac{\sqrt{\alpha}}{2}(\tau_1-\tau_2)\biggr)\Biggr).   \nonumber
\ea
After straightforward calculations of r.h.s. in (\ref{BB}), of $B$ in (\ref{BG}), and finally, of the potentials $V^{(2)}$ in (\ref{V2}) and $V^{(1)}$ in (\ref{V}), one obtains:
\ba
&&C_1C_2\biggl[2\alpha +2C_1^{\prime}C_2^{\prime}+2\alpha (y_1C_2+y_2C_1)+\alpha_1C_2+\alpha_2C_1\biggr]=4\alpha^2\biggl[\varphi_1 + (\varphi_1+\varphi_2)\varphi_1^{\prime}\biggr];
\nonumber\\
&&B=\frac{2\varphi_1^{\prime}+\alpha (\varphi_1^2+2\varphi_1\varphi_2)+F(\tau_2)}{\varphi_1+\varphi_2};
\nonumber\\
&&V^{(2)}=\frac{2\varphi_1^{\prime}-\alpha \varphi_1^2+F(\tau_2)}{\varphi_1+\varphi_2};
\nonumber\\
&&V^{(1)}=V^{(2)}-2(C_1^{\prime}+C_2^{\prime})=\frac{-2\varphi_1^{\prime}-\alpha \varphi_1^2+F(\tau_2)}{\varphi_1+\varphi_2},  \nonumber
\ea
where the functions $\varphi_1,\,\varphi_2$ were introduced for compactness:
\ba
&&\varphi_1(\tau_1)\equiv \frac{1}{4\gamma_1\gamma_2}\biggl(\gamma_1^2\gamma_2^2\exp(\sqrt{\alpha}\tau_1)+\omega_1^2\omega_2^2\exp(-\sqrt{\alpha}\tau_1)\biggr);
\nonumber\\
&&\varphi_2(\tau_2)\equiv \frac{1}{4\gamma_1\gamma_2}\biggl(\gamma_1^2\omega_2^2\exp(\sqrt{\alpha}\tau_2)+\omega_1^2\gamma_2^2\exp(-\sqrt{\alpha}\tau_2)\biggr).\nonumber
\ea
In variables $\tau_1,\,\tau_2,$ the kinetic term in Minkowski space includes the same factor $(\varphi_1+\varphi_2)^{-1}$ as the potentials $V^{(1)},\,V^{(2)}$:
\begin{equation}
\partial_1^2-\partial_2^2=\frac{4}{\alpha (\varphi_1+\varphi_2)}(\partial_{\tau_1}^2-\partial_{\tau_2}^2)  \nonumber
\end{equation}
and therefore, the corresponding Schr\"odinger equations are amenable to the so-called $R-$separation of variables \cite{miller}, i.e. the two-dimensional problems
are reduced to a pairs of one-dimensional problems. The property of $R-$separation is confirmed by existence of the symmetry operator of second order in derivatives:
\begin{equation}
\widetilde{R}^{(1)}\equiv Q^+Q^- - (H^{(1)})^2 -2\widetilde\beta H^{(1)} = \frac{-4}{\varphi_1+\varphi_2}(\varphi_1\partial_{\tau_2}^2+\varphi_2\partial_{\tau_1}^2) +
\frac{\varphi_2(\varphi_1^2+2\alpha \varphi_1') +\alpha\varphi_1F}{\varphi_1 + \varphi_2} + \frac{\widetilde\beta^2}{4},  \nonumber
\end{equation}
which depends explicitly on the value of parameter $\widetilde\beta$ chosen in (\ref{V2}).

\section{Nondiagonalizable $g_{nm}$}

In the previous Section the case of diagonalizable constant "metric" (\ref{g}) in the supercharge was considered in a general form.
It is also interesting to consider the case of nondiagonalizable matrix $g_{nm}$, that will be done in the present Section. Thus, let us take:
\begin{equation}\label{nondiag}
(g_{11}+g_{22})^2 \leq 4g_{12}^2;\quad g_{11}g_{22}g_{12}\neq 0.
\end{equation}
Starting similarly to the beginning of Subsection 3.1, from (\ref{1}) one has:
\be
\partial_1\widetilde C_1=g_{11}V; \quad
\partial_2\widetilde C_2=-g_{22}V; \quad
\partial_1\widetilde C_2-\partial_2\widetilde C_1=2g_{12}V,   \nonumber
\ee
and therefore, functions $\widetilde C_1, \,\widetilde C_2,\, V$ can be expressed in terms of one function:
\begin{equation}\label{C}
\widetilde C_1=\frac{1}{g_{22}}\partial_2 C;\quad \widetilde C_2=-\frac{1}{g_{11}}\partial_1 C;\quad V=\frac{1}{g_{11}g_{22}}\partial_1\partial_2 C,
\end{equation}
which satisfies the second order differential equation:
\begin{equation}\label{eq}
(g_{22}\partial_1^2+g_{11}\partial_2^2+2g_{12}\partial_1\partial_2)C=0.
\end{equation}
By choosing normalization of "metric" with
$$g_{11}\equiv 1$$
and introducing the real combination:
\begin{equation}
\gamma\equiv g_{12}+ \sqrt{g_{12}^2-g_{22}},  \nonumber
\end{equation}
the equation (\ref{eq}) becomes:
\begin{equation}
(\gamma\partial_1+\partial_2)(g_{22}\partial_1+\gamma\partial_2)C=0.   \nonumber
\end{equation}
Its general solution can be written as:
\begin{equation}
C(\vec x)=C_1(y_1)+C_2(y_2),       \nonumber
\end{equation}
with new variables:
\begin{equation}
y_1\equiv \gamma x_2-x_1;\quad y_2\equiv \gamma x_1-g_{22}x_2.  \nonumber
\end{equation}
Now, the initial functions are:
\be
\widetilde C_1 =\omega C_1^{\prime}(y_1)-C_2^{\prime}(y_2); \quad
  \widetilde C_2 =C_1^{\prime}(y_1)-\gamma C_2^{\prime}(y_2); \quad
  V = -\biggl(\omega C_1^{\prime\prime}(y_1)+\gamma C_2^{\prime\prime}(y_2)\biggr),   \nonumber
\ee
where the constants
\begin{equation}
\omega\equiv \frac{\gamma}{g_{22}};\quad 2g_{12}=\gamma + \frac{1}{\omega},    \nonumber
\end{equation}
and the condition (\ref{nondiag}) for $\omega\gamma\neq 1$ reads:
\begin{equation}
(1-\gamma^2)(1-\frac{1}{\omega^2})\leq 0.   \nonumber
\end{equation}

It follows from (\ref{2}) that:
\begin{eqnarray}
  &&(\partial_1^2-\partial_2^2)\widetilde C_1 +2(\partial_1B)-2V\widetilde C_1-2\biggl(\partial_1V^{(2)}+g_{12}\partial_2V^{(2)}\biggr)=0; \label{C1C2} \\
  &&(\partial_1^2-\partial_2^2)\widetilde C_2 -2(\partial_2B)-2V\widetilde C_2-2\biggl(g_{22}\partial_2V^{(2)}+g_{12}\partial_1V^{(2)}\biggr)=0. \label{C2C1}
\end{eqnarray}
One can take the sum of derivatives of (\ref{C1C2}) and (\ref{C2C1}) over $x_2$ and $x_1,$ respectively, to obtain:
\ba
&&(\partial_1^2-\partial_2^2)(\partial_2\widetilde C_1+\partial_1\widetilde C_2)-2\partial_2(V\widetilde C_1)-2\partial_1(V\widetilde C_2)-2\biggl(\partial_1\partial_2V^{(2)}+g_{12}\partial^2_2V^{(2)}\biggr)-\nonumber\\
&&-2\biggl(g_{22}\partial_1\partial_2V^{(2)}+g_{12}\partial^2_1V^{(2)}\biggr)=0.    \nonumber
\ea
After straightforward calculations, for the case $\gamma\omega\neq 1,$ one has equation for potential $V^{(2)}:$
\ba
&&\omega(1-\gamma^2)C_1^{IV}(y_1)+\omega^2\biggl((C^{\prime}_1(y_1))^2\biggr)^{\prime\prime}-\gamma^3(1-\frac{1}{\omega^2})C^{IV}_2(y_2)-
\gamma^2\biggl((C^{\prime}_2(y_2))^2\biggr)^{\prime\prime}+\nonumber\\
&&+(1-\gamma^2)\partial_{y_1}^2V^{(2)}-\gamma^2(1-\frac{1}{\omega^2})\partial_{y_2}^2V^{(2)}=0.    \nonumber
\ea
Its general solution is:
\be
V^{(2)}=-\omega C_1^{\prime\prime}-\frac{\omega^2}{1-\gamma^2}(C^{\prime\prime}_1)^2-\gamma C^{\prime\prime}_2-\frac{\omega^2}{\omega^2-1}(C^{\prime}_2)^2+\widetilde V^{(2)}, \label{tilde}
\ee
where $\widetilde V^{(2)}$ is the general solution of the homogeneous equation:
\begin{equation}\label{homo}
(\partial_{y_1}^2-\Omega^2\partial_{y_2}^2)\widetilde V^{(2)}=0; \quad \Omega\equiv |\frac{\gamma}{\omega}|\sqrt{\frac{1-\omega^2}{1-\gamma^2}}
\end{equation}
Equation (\ref{homo}) is easily solved in terms of complex conjugate variables:
\begin{equation}\label{V222}
\widetilde V^{(2)}=G(z)+\bar G(\bar z); \quad z\equiv y_2+i\Omega y_1; \quad \bar z\equiv y_2-i\Omega y_1
\end{equation}
with arbitrary function $G$ and its conjugate $\bar G.$

Another linear combination of equations (\ref{C1C2}) and (\ref{C2C1}), with coefficients $\gamma$ and $-1,$ respectively, gives also the equation for potential $V^{(2)}$:
\ba
&&(\partial_1^2-\partial_2^2)(\gamma \widetilde C_1 - \widetilde C_2)+2(\gamma^2-g_{22})\partial_{y_2}B-\nonumber\\
&&-2V(\gamma \widetilde C_1 - \widetilde C_2)
-2(\gamma \partial_1+\gamma g_{12}\partial_2-g_{22}\partial_2-g_{12}\partial_1)V^{(2)}=0,     \nonumber
\ea
which for $\omega\gamma\neq 1$ and $V^{(2)}$ from (\ref{tilde}), (\ref{V222}), means that:
\begin{equation}\label{L1}
2B-(1-\frac{\gamma}{\omega})V^{(2)}+2\omega C^{\prime}_1C^{\prime}_2+\frac{i(1-\gamma^2)\Omega}{\gamma}(G-\bar G)=L_1(y_1),
\end{equation}
where $L_1$ is an arbitrary function of its argument. To check the equivalence of obtained results for $V^{(2)}$ and $B$ to initial equations (\ref{C1C2}), (\ref{C2C1}), one
more linear combination of them, with coefficients $g_{22}$ and $-\gamma$ can be calculated. The result is:
\begin{equation}\label{L2}
2B-(1-\frac{\gamma}{\omega})V^{(2)}+2\omega C^{\prime}_1C^{\prime}_2+\frac{i(1-\gamma^2)\Omega}{\gamma}(G-\bar G)=L_2(y_2),
\end{equation}
also with arbitrary function $L_2.$ This means that $L_1(y_1)=L_2(y_2)\equiv e = const$ both in (\ref{L1}), and in (\ref{L2}).

The last equation that must be solved in the considered case of nondiagonalizable $g_{nm}$ is the equation (\ref{3}). Using derivatives of (\ref{2}), one obtains from (\ref{3}):
\begin{equation}\label{2--3}
\partial_p(V\widetilde C_p)-\frac{1}{2}(\partial_1^2-\partial_2^2)(\partial_p\widetilde C_p)-2VB-\widetilde C_p\partial_pV^{(2)}=0.
\end{equation}
By the direct calculations it can be transformed to the following form:
\begin{eqnarray}
&&-\partial_{y_1}\Biggl[C^{\prime}_1\biggl((\gamma - \omega)(G+\bar G)-e\omega + \frac{i(1-\gamma^2)\Omega\omega}{\gamma}(G-\bar G)\biggr)+\frac{(\gamma -\omega)(1-\gamma^2)}{2}C_1^{\prime\prime\prime}-\nonumber\\
&&-(\gamma -\omega)\omega^2C^{\prime}_1\biggl(\frac{(C^{\prime}_1)^2}{1-\gamma^2}+\frac{(C^{\prime}_2)^2}{\omega^2-1}\biggr)\Biggr]= \nonumber\\
&& =\frac{\gamma}{\omega}\partial_{y_2}\Biggl[C^{\prime}_2\biggl((\gamma - \omega)(G+\bar G)-e\omega + \frac{i(1-\gamma^2)\Omega\omega}{\gamma}(G-\bar G)\biggr)+\frac{(\gamma -\omega)\gamma^2(\omega^2-1)}{2\omega^2}C_2^{\prime\prime\prime}-\nonumber\\
&&-(\gamma -\omega)\omega^2C^{\prime}_2\biggl(\frac{(C^{\prime}_2)^2}{\omega^2-1}+\frac{(C^{\prime}_1)^2}{1-\gamma^2}\biggr)\Biggr]. \label{20-20}
\end{eqnarray}

Equation (\ref{20-20}) has the functional-differential form, and again, no regular algorithm to solve it exists. We can only rely on luck choosing some specific values of parameters. For example, the initial problem can be simplified essentially for the coinciding values:
\begin{equation}
\omega =\gamma\neq 1;\quad g_{22}=1;\quad 2g_{12}=\gamma + 1/\gamma ; \quad \Omega =1,   \nonumber
\end{equation}
for which many terms in (\ref{20-20}) vanish and it takes much simpler form:
\begin{equation}\label{partial}
\partial_{y_1}[C^{\prime}_1(y_1)\Phi(\vec y)] = - \partial_{y_2}[C^{\prime}_2(y_2)\Phi(\vec y)],
\end{equation}
with
\begin{equation}
\Phi(\vec y)\equiv G(z)-\overline{G}(\bar z); \quad z=y_2+iy_1;\quad \bar z=y_2-iy_1;\quad y_1 = \gamma x_2-x_1;\quad y_2 = \gamma x_1-x_2,   \nonumber
\end{equation}
(function $G(z)$ was shifted here by a constant).

Again, similarly to analysis of Subsection 3.1, we may use the results obtained for the same equation as (\ref{partial}) (up to change in notations of functions and to a suitable change of coordinates) in papers \cite{david} and \cite{bin} in another context - of Euclidean plane.
Specifically, the function $C_1^{\prime}(y_1)$ must be identified with function $C_+(y_1)$ in paper \cite{bin}, the function $C_2^{\prime}(y_2)$ with
$-iC_-(iy_2)$, and $\Phi(\vec y)$ with the old $F(\vec y).$ Similarly to Eq.(\ref{reality}), the additional condition of reality must be fulfilled:
\begin{equation}\label{real}
C_+(y_1)=\overline{C_+}(y_1);\quad C_-(y_2)=-\overline{C_-}(-iy_2);\quad F(y_1, iy_2)=-\overline{F}(y_1, -iy_2).
\end{equation}

The list of solutions of (\ref{partial}) which satisfy the conditions (\ref{real}) is rather long, but most part of them have infinitely many singularities and/or nonphysical asymptotic behavior along some directions on the plane.
The detailed analysis gives the only pair of partner potentials which after a suitable choice of parameters are free from the above disadvantages:
\ba
V^{(1,2)}=\mp \lambda\gamma\mu \frac{\sinh(\lambda y_1)}{\cosh^2(\lambda y_1)}-
\frac{\gamma^3\mu^2}{1-\gamma^2}\frac{1}{\cosh^2(\lambda y_1)}+2\gamma\sinh(\lambda y_1)\cos(\lambda y_2). \label{28}
\ea
For illustration, the plot of $V^{(1)}(x_1,x_2)$ with parameters $\lambda =1,\, \gamma =0,2,\, \mu =0,1$ is given at Figure 2 in variables $x_1, x_2.$

\vspace{6pt}

\begin{center}
\includegraphics[height=5cm]{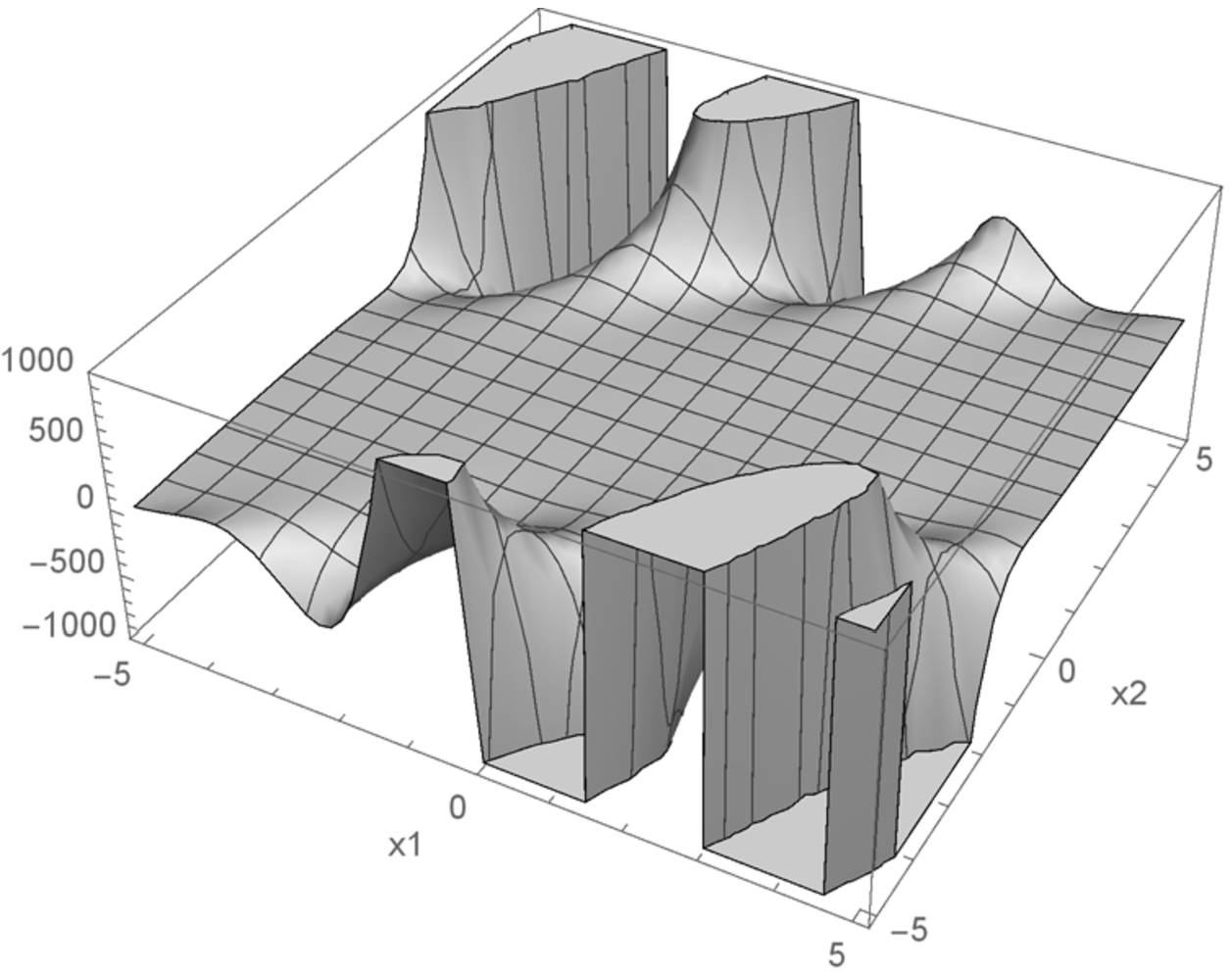}

\noindent{\it Fig.2} Plot of the potential $V^{(1)}(x_1,x_2)$ (\ref{28}) for $\lambda =1,\, \gamma =0,2,\, \mu =0,1.$

\end{center}

\vspace{4pt}
Up to now, we restricted ourselves by the choice $\omega =\gamma \neq 1$ . Thus, the case
\begin{equation}
\gamma = \omega =1;\quad g_{11}=g_{22}=g_{12}=1;\quad \Omega =1, \label{omega-gamma}
\end{equation}
has to be considered separately. According to (\ref{eq}),
\begin{equation}\label{eqq}
(\partial_1+\partial_2)^2C(\vec x)=0,
\end{equation}
i.e., in "light cone" coordinates $x_{\pm}\equiv x_1\pm x_2,$
\begin{equation}\label{CC}
C(\vec x)=x_+C_-(x_-)+A_-(x_-),
\end{equation}
with arbitrary functions $C_-$ and $A_-.$ Substitution into (\ref{C}) and (\ref{C1C2}), (\ref{C2C1}) gives:
\begin{eqnarray}
&&-2C_-''+\partial_-B+V(x_+C_-'+A_-')-\partial_+V^{(2)}=0; \label{CB} \\
&&\partial_+B-VC_-=0. \label{BC}
\end{eqnarray}
Therefore,
\ba
&&B=-\frac{1}{2}x_+^2C_-''C_--x_+A_-''C_-+B_-(x_-); \label{BBBB}\\
&&V^{(2)}=-\frac{x_+^3}{12}(C_-^2)'''-\frac{x_+^2}{2}(A_-'C_-)''+(B_-'-A_-''A_-'-2C_-'')x_++D_-(x_-), \label{VVVV}
\ea
with arbitrary $B_-,\, D_-.$ From the asymptotic behaviour of $V^{(2)}$ at $x_+\to\infty ,$ it follows that
\begin{equation}\label{CCCC}
 (C_-^2)'''=0.
\end{equation}

From (\ref{2--3}), one can derive the relation:
\begin{equation}\label{222a}
\widetilde C_p\partial_p(V-V^{(2)})-2VB=0,
\end{equation}
which leads in particular to the necessary condition onto function $C_-:$
\begin{equation}\label{222b}
-\frac{1}{2}C_-'(C_-^2)'''+\frac{1}{6}(C_-^2)^{IV}C_--(C_-'')^2C_-=0
\end{equation}
Thus, only two options are possible for $C_-$: either $C_-=cx_-$ or $C_-=const.$
One can check that both cases lead to the potentials $V^{(2)}$ with unphysical asymptotic behaviour.

\section{Conclusions}

It might be instructive to point out the situations when one of partner potentials $V^{(2)}(\vec x)$ is trivial, namely equal to constant. In such a case, the spectrum of $H^{(2)}$ is continuous, and its eigenfunctions $\Psi^{(2)}$ are plane waves. Nevertheless, due to intertwining relations (\ref{intertw}) the discrete spectrum of its partner $H^{(1)}$ may exist, including the eigenvalues which correspond to linear combinations of normalizable zero modes of the supercharge $Q^-$ (see \cite{new}). To obtain the result analytically, the nontrivial problem $Q^-\Omega_n(\vec x)=0$ must be solved. One such opportunity can be provided by the separability of variables in operator $Q^-$ and solvability of obtained one-dimensional problems. For Euclidean space the analogous procedure was formulated and realized in papers \cite{new}, \cite{ioffe-reviews}, \cite{ai}: it was called as SUSY separation of variables I. In our present case of Minkowski space, one can check that $Q^{\pm}$ of the general form (\ref{charge}) allow the separation of variables by means of a specially adopted similarity transformation:
\ba
q^-&=&\exp{(+\chi)} Q^+ \exp{(-\chi)}=\partial_1^2+\partial_2^2+F_+(2x_1)-F_-(2ix_2); \label{c1} \\
\chi (\vec x)&=&-\frac{1}{2}\biggl(\int C_1(y_1)dy_1+\int C_2(y_2)dy_2\biggr). \label{c2}
\ea
Finally, the solvability of the two-dimensional problem depends on solvability of one-dimensional problems with "potentials" $F_{\pm}.$

Among the constructed potentials, several examples can be investigated by this algorithm after a suitable choice of parameters. The simplest pair is given by (\ref{2222222}) with $b=-2,\, n=m=0,$ for which the potential $V^{(2)}=0$ and its partner in terms of polar coordinates $x_1=r \cos\phi,\, x_2=r \sin\phi$ is:
\be
V^{(1)}(r, \phi)= -8\frac{x_1^2-x_2^2}{(x_1^2+x_2^2)^2}= -16\frac{\cos{2\phi}}{r^2}. \label{c3}
\ee
The Schr\"odinger equation with potential (\ref{c3}) can be solved by the standard procedure of $R$-separation of variables. By the way, the functions $F_{\pm}$ vanish in this example.

The next examples are (\ref{44}) and (\ref{44444}), both with $b=c=-\lambda ,\, a=k=0,$ the potentials (\ref{55}) with $k=-\lambda,\, k_1=k_2=0$
and potentials (\ref{555}) with $k=+\lambda,\, k_1=k_2=0.$ The described algorithm works for these models, for all of them the functions $F_{\pm}$ vanish.
As formulated above, this means an absence of discrete spectrum for Hamiltonians $H^{(2)}.$ Nevertheless, the wave functions of the partner Hamiltonian $H^{(1)}$ can be
built as linear combinations of zero modes of (\ref{c1}), multiplied by $\exp{(-\chi)}$ with $\chi$ given by (\ref{c2}). One can check that this exponential multiplier is decreasing for all potentials (\ref{44}), (\ref{44444}), (\ref{55}), (\ref{555}), leading to a variety of normalizable eigenfunctions.

The SUSY approach in Quantum Mechanics was generalized onto the two-dimensional quantum models in Minkowski space. Polynomial deformation of SUSY algebra was considered with supercharges of second order in derivatives. It is known that in this case the components $H^{(1)},\,H^{(2)}$ of Superhamiltonian correspond to completely integrable models with symmetry operators of fourth order in momenta. In the components, intertwining relations are equivalent to the systems of nonlinear differential equations which can not be solved in the most general form. Here, it was solved separately for different forms of constant "metric" matrix in second order part of the supercharges. The list of corresponding potentials both for diagonalizable and nondiagonalizable matrices $g_{nm}$ was obtained. Some of these potentials are too singular and/or have unlimited decrease  asymptotic behaviour at infinity, but others have reasonable analytic properties. It was shown that several potentials obey the properties of shape invariance and are amenable to peculiar procedure of SUSY separation of variables.

\section{Acknowledgments}

The work was partially supported by the grant of Saint Petersburg State University N11.38.223.2015 (E.V.K.).

\end{document}